 \documentclass[aps,prl,twocolumn,superscriptaddress,showpacs,floatfix]{revtex4}

\usepackage{graphicx}
\usepackage{bm}

\begin{document}


 \title{Thermodynamics Predicts How Confinement Modifies Hard-Sphere Dynamics}


 \author{Jeetain Mittal}
 \affiliation{Department of Chemical Engineering, The University of Texas at 
 Austin, Austin, TX}

 \author{Jeffrey R. Errington}
 \affiliation{Department of Chemical and Biological Engineering, University at 
 Buffalo, The State University of New York, Buffalo, NY}

 \author{Thomas M. Truskett$^{1,}$}
 \affiliation{Institute for Theoretical Chemistry, The University of Texas at Austin, Austin, TX}


 \date{\today}

\begin{abstract}
We study how confining the equilibrium hard-sphere fluid to
restrictive one- and two-dimensional channels with smooth interacting
walls modifies its structure, dynamics, and entropy using molecular
dynamics and transition-matrix Monte Carlo simulations.  Although
confinement strongly affects local structuring, the relationships
between self-diffusivity, excess entropy, and average fluid
density are, to an excellent approximation, independent of channel
width or particle-wall interactions.  Thus, thermodynamics can be
used to predict how confinement impacts dynamics.
\end{abstract}
\maketitle
%
%
The molecular dynamics of fluids confined to small spaces
can differ significantly from the bulk.  These differences have generated 
wide interest because confined fluids feature prominantly
in both nature and technology.  Examples include 
dynamics of water near proteins or in concentrated cellular
environments, transport processes across biological membranes, and
fluid flows encountered in micro- or nanofluidic devices, to mention
a few.
Given that a significant fraction of the molecules in
these systems populate highly inhomogeneous interfacial environments,
it is easy to
appreciate why confinement has nontrivial consequences for their
transport coefficients (e.g., diffusivity and viscosity). 
Nonetheless, a theoretical framework that can reliably predict these 
consequences has been slow to develop.

One logical starting point is to ask whether confinement induced 
modifications to equilibrium fluid properties, such as the density, 
can explain some of the observed differences in 
dynamics~\cite{reiter,alba1,curro}.  
For instance, if the presence of a
strongly attractive substrate increases the local fluid density near 
the fluid-substrate interface, 
one might naturally expect a corresponding decrease in particle mobility
near that interface, and vice versa.  
This type of argument is physically intuitive, and it 
has been recently used to 
rationalize why nanoscale materials exhibit glass transition
temperatures
that are shifted relative to their bulk values~\cite{curro}.  
However, it seems doubtful that average structural quantities alone can 
account for
the wide variety of dynamical behaviors observed in both simulations
and experiments of confined fluids~\cite{jones,binder2,lowen_ss}.  
As a result, it is natural to ask whether 
other equilibrium measures, such as the entropy, can provide
additional insights.  Unfortunately, it is currently difficult to
obtain the necessary experimental data for testing 
these possible connections between thermodynamics and 
dynamics for confined fluids.
Thus, simulation results on simple and well-defined model systems are of great 
complementary value.

In this Letter, we advance the current understanding of 
the relation between thermodynamics and dynamics in inhomogeneous systems
by resolving the following two questions. 
(i) Can either the entropy or the average density be used to determine the extent 
to which confinement alters the diffusivity of a hard-sphere (HS) fluid?
(ii) If so, do the specific interactions between particles and 
the channel boundaries significantly impact the result?   
While the confined HS system represents arguably the most 
elementary and well-studied model 
for inhomogeneous colloidal and molecular fluids, 
there is still
surprisingly little known about the possible connections between its basic
thermodynamic and kinetic properties.  If such connections do exist
and prove to be robust,
it suggests that equilibrium theories for inhomogeneous fluids
might generally provide important information regarding how
confinement 
modifies 
the transport properties of fluids.

One reason to speculate that entropy could be a reliable predictor
for how confinement affects the diffusivity is its empirical 
success for
capturing the dynamical behavior of bulk materials.  
In particular, computer simulation studies
have demonstrated that the single-component HS
and Lennard-Jones fluids, along with a variety of models for liquid metals,
exhibit, to a very good approximation, a one-to-one relationship
between diffusivity and excess 
entropy over a broad range of thermodynamic 
conditions~\cite{rosen1,dzu1,rosen2,bret,bast}.  
Excess
entropy has also been shown to accurately 
capture the behavior of diffusion phenomena
in fluid mixtures~\cite{sam1,sam2,hoyt} as well as those in 
solid-state ionic conductor and quasicrystalline materials~\cite{dzu1}. 
 
To explain the origin of the correspondence between excess entropy
and diffusivity in bulk materials, several researchers
have presented independent derivations of apparent scaling laws
relating the two quantities.  
The earliest that we are aware of is due
to Rosenfeld and is motivated by a variational 
thermodynamic perturbation theory~\cite{rosen1,rosen2}.  
Dzugutov later used arguments based on kinetic
theory to justify a similar scaling~\cite{dzu1}, and recently
mode-coupling theory has been employed to establish an approximate
basis for the observed connection~\cite{sam1,sam2}. 
Despite the large amount of effort that has been devoted to
justifying these scaling laws theoretically and testing their validity
for bulk materials, to our knowledge the relationship between excess
entropy and diffusivity has never been tested in inhomogeneous
fluids, nor has it been used as a tool 
to understand how confinement affects dynamics.

To carry out such a test, we studied how the structure, 
thermodynamics, and dynamics of the single-component HS fluid
confined to restrictive 
two-dimensional (2$d$) or one-dimensional (1$d$) channels 
(``films'' or ``pores'', respectively) 
bounded by smooth, interacting walls differ from those 
of the bulk system.  
We considered five different 2$d$ channel sizes that were effectively 
macroscopic in the $x$ and $y$ directions and had particle-center accessible 
dimensions in the confining $z$ direction of $h_z=15$, 10, 7.5, 5, and
2.5, repectively.  
We also considered 
three 1$d$ channel sizes that were effectively 
macroscopic in the $x$ direction and
had particle-center accessible dimensions in the confining $y$ and
$z$ directions of $h_y\times h_z= 7.5 \times 7.5$, $7.5 \times 5$, 
and $5 \times 5$, respectively.   
To simplify the notation, we have implicitly 
non-dimensionalized all lengths in this study 
by the HS particle diameter $\sigma$
and all times by the combination $\sigma \sqrt{m \beta}$, 
where $m$ is particle mass, $\beta=1/k_{\mathrm B}T$, $k_{\mathrm B}$
is Boltzmann's constant, and $T$ is temperature.  Consequently,
all energies are given per unit $k_{\mathrm B}T$.  
Position-dependent interations between the particles and 
the confining channel walls $u(\zeta)$ were calculated 
using a square-well potential:
\begin{eqnarray}
u(\zeta) &=&\infty~~~~~~~~~~~~~~~~~~~~~\zeta < 1/2 \nonumber \\
     &=&\epsilon_{\mathrm w}~~~~~~~~~~~~1/2 \leq \zeta < 1 \nonumber \\
     &=&0~~~~~~~~~~~~~~~~~~~~~~\zeta \geq 1,
\label{wall_int}
\end{eqnarray}
where $\zeta$ represents the shortest distance between a given
particle center and the wall of interest, 
and $\epsilon_{\mathrm w}$ is the strength of the 
effective particle-wall interaction.  We 
considered five specific particle-wall interactions for the
2$d$ channels:  $\epsilon_{\mathrm w}=1$ and $\epsilon_{\mathrm
  w}=0.5$ representing repulsive walls, $\epsilon_{\mathrm w}=0$
representing ``hard'' but neutral walls, and $\epsilon_{\mathrm
  w}=-0.5$ and $\epsilon_{\mathrm w}=-1$ 
representing attractive walls.  Only the hard walls were
considered for the three highly restrictive 1$d$ channels.  

To monitor kinetic processes 
in these various systems, we performed a series of 
event-driven
molecular dynamics simulations~\cite{rap} in the microcanonical
ensemble using $N=4500$ hard spheres.   
Periodic boundary conditions were employed in the $d$
``free'' directions (i.e., directions not confined by walls).  
The dimensions of the simulation cell in the
periodic directions were set to various values to simulate fluids with 
different average number densities that span the stable 
equilibrium range,
from the dilute gas to the
fluid at its freezing transition.
We extracted the self-diffusivity of the fluid $D$ 
by fitting the long-time ($t \gg 1$) behavior of the 
average mean-squared 
displacement of the particles 
to the Einstein relation $<\Delta {\bf r}_d^2> = 2dDt$, where
$\Delta {\bf r}_d^2$ corresponds to the mean-square displacement 
in the $d$ periodic directions.      
We also calculated $D$ for several state
points with both smaller ($N=3000$) and
larger ($N=6000$) particle numbers to verify that 
system size effects in the periodic directions were negligible.

We determined the behavior of the 
excess entropy per particle $s^{\mathrm{ex}}$ using grand canonical 
transition-matrix Monte Carlo (GC-TMMC) simulations~\cite{jeff1}. 
Here, $s^{\mathrm{ex}}$ is defined to be the difference between the 
entropy per particle of the fluid and that of an ideal gas with the
same spatial distribution of the particle number density.  
GC-TMMC simulations require fixed values for the activity
$\xi$~\cite{note1}, 
the particle-center accessible dimensions \{$h_x, h_y, h_z$\} that
define the volume of the simulation cell $V=h_x h_y h_z$, and 
the reciprocal temperature $\beta$. 
For all simulations conducted here, we set $\xi = 1$, $\beta = 1$,
and we used the particle-wall interactions given by Eq.~(\ref{wall_int}).
The values of 
$h_y \times h_z$ or $h_z$ are determined by the confining
dimensions of the 1$d$ or 2$d$ channels, respectively, and the 
remaining periodic dimension(s) were chosen to satisfy $V=1000$.
Indistinguishable results were
obtained for systems of size $V=500$.

The key quantities extracted from the GC-TMMC simulations were the 
total particle 
number probability distribution $\Pi(N)$, 
the excess configurational energy 
$U^{\mathrm{ex}}(N)$, and the $N$-specific 
spatial density distribution $\rho(N,{\bf r})$, each evaluated over a 
range of particle numbers spanning from $N = 0$ to $N = 984$.  
Using basic 
arguments from statistical mechanics~\cite{azp00,htdavis}, one can relate 
these quantities to $s^{\mathrm{ex}}$ for the inhomogeneous 
HS fluid:
\begin{eqnarray}
s^{\mathrm{ex}}(N)/k_{\mathrm B} = N^{-1} \{\ln [\Pi(N)/\Pi(0)]
  -N{\mathrm {ln}}{\xi} + \ln N! \nonumber \\ 
- N\ln N + \beta U^{\mathrm{ex}}(N) +\int \rho(N,{\bf r}) \ln
\rho(N,{\bf r}) d {\bf r} \}
\label{s1}
\end{eqnarray}
Given that $V=1000$ is fixed, Eq.~(\ref{s1})
provides $s^{\mathrm{ex}} (\rho_h)$ within the 
range $0 \leq \rho_h \leq 0.984$, where $\rho_h = N/V$ is the number
density based on the particle-center accessible volume. 

\begin{figure}
\scalebox{.9}{\includegraphics{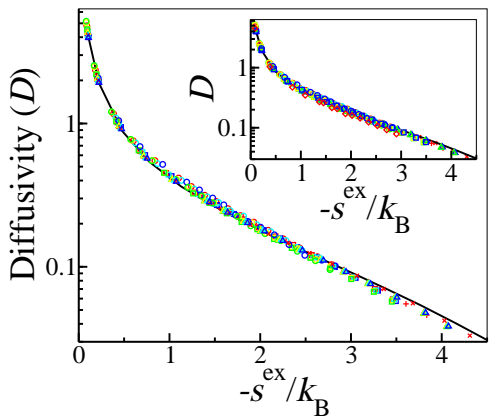}}
\caption{\label{D_entropy}Self-diffusivity $D$ versus the negative 
of excess entropy per particle $-s^{{\text {ex}}}/k_{\mathrm B}$
for the bulk HS fluid (solid curve) and for the HS fluid 
in 2$d$ channels (symbols).  The symbols correspond to 
$h_z=2.5$ (\text{circle}), 5 (\text{square}), 7.5 (\text{plus}), 
10 (\text{triangle up}) and 15 (\text{cross}).  
The color codes are $\epsilon_w=$1 (\text{blue}), 0.5 (\text{cyan}), 
0 (\text{red}), -0.5 (\text{yellow}), and -1 (\text{green}). 
Inset is the same plot with added points for the 1$d$ channels:
$5\times5$, $7.5\times7.5$, and $5\times7.5$ shown by red diamonds.}  
\end{figure}
First, we discuss our observations for the utility of excess entropy 
$s^{{\text{ex}}}$ in predicting how confinement affects
dynamics.  Fig.~\ref{D_entropy} shows a parametric plot of
$D$ versus $-s^{{\text{ex}}}$ for the HS fluid both in the bulk
and confined to the 17 different 2$d$ channels 
($h_z=2.5$, 5, 10 with $\epsilon_{\mathrm w}=1$, 0.5, 0, 
-0.5, -1 
and $h_z=7.5$, 15 with $\epsilon_{\mathrm w}=0$).  
The data, which encompasses the
dynamic behavior of the equilibrium fluid from the dilute gas 
to the freezing transition, spans three decades in $D$.
The collapse of the data onto a single master curve indicates that, 
to an excellent approximation, the simple one-to-one
correspondence between $D$ and $s^{{\text{ex}}}$ for the bulk 
HS fluid also holds when the fluid is severely confined. 
Moreover, the quality of the data collapse is largely independent of 
either channel width (including ``particle scale'' channels with $h_z=2.5$)
and the sign or magnitude of the particle-wall interaction.  Data for
the HS fluid confined to the 3 rectanglular 1$d$ channels described
above are superimposed in the inset of Fig.~\ref{D_entropy}.  As
can be seen, they also approximately
collapse onto the bulk HS relationship between $D$ and $s^{\text{ex}}$. 
In fact the diffusivities for approximately 50, 70, and 90$\%$ of the 
confined state points shown in 
Fig.~\ref{D_entropy} are within 3, 5, and 10 $\%$ 
respectively of the bulk value at the same excess entropy.

Having established that $s^{\text{ex}}$, an equilibrium thermodynamic
property, can be used to predict how various confining
environments impact the self-diffusivity of the HS system, 
it is natural to press forward and ask whether the same 
predictive information is contained
in an even simpler structural measure:  the average density.
\begin{figure}
\scalebox{.9}{\includegraphics{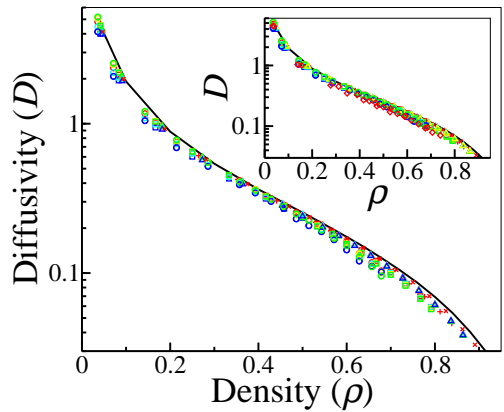}}
\caption{\label{D_density}Self-diffusivity $D$ vs average density 
$\rho= N/V'$ based on the the total system volume $V'$. 
Systems shown include the bulk 
HS fluid (solid curve) and the HS fluid confined to 
the 2$d$ channels (symbols) discussed in the text.  
The inset includes the data from the HS fluid
confined to the 1$d$ channels.  Symbols for the main panel and the inset 
are identical to those given in Fig.~\ref{D_entropy}.}
\end{figure}
Here, one needs to be specific because there are two different
definitions of average density that are commonly
used to characterize
inhomogenenous HS fluids ($\rho_h=N/V$ and $\rho=N/V'$).  
The former is based on the
particle-center-accessible volume $V=h_x h_y h_z$, while the latter 
is based on the total system volume; i.e., $V'=V$ for bulk systems, 
$V'=h_x h_y (h_z+1)$ for 2$d$
channels, and $V'=h_x (h_y+1) (h_z+1)$ for 1$d$ channels.  Schmidt and 
L\"{o}wen~\cite{lowen2dp} and, later, Zangi and Rice~\cite{rice2dp}
demonstrated that $\rho$ is in fact the
relevant density for the lateral (i.e., periodic) component(s) of the 
pressure, and thus $\rho$ is also a natural independent variable for the 
free energy of the system.  Below, we investigate
whether $\rho$ is also an accurate predictor for how confinement 
effects HS diffusivity.
         
Fig.~\ref{D_density} shows $D$ as a function of $\rho$ for the bulk 
HS fluid as well as for the HS fluid confined to 
the 2$d$ and 1$d$ channels.  
The collapse of the data, while not perfect, 
demonstrates a very strong correlation between 
$D$ and $\rho$ that is nearly independent of the confining dimensions and 
the particle-wall interactions.  
Approximately 70 and 90$\%$ of the diffusivities for the confined 
state points shown in 
Fig.~\ref{D_density} are within 10 and 20$\%$, 
respectively, of the bulk value at the same average density.
This is another
significant result because, unlike $s^{\text{ex}}$, $\rho$ is
intuitively simple to understand and trivial to 
determine in simulations (e.g.,
it is specified in microcanonical and canonical simulations). 
As is expected, the systems that exhibit the most noticeable deviation
from bulk behavior in Fig.~\ref{D_density}
are the ones for which the fluid is under the most severe 
confinement, i.e., channels with dimensions 
comparable to the particle diameter. In these cases, it appears
that specific fluid structuring (e.g., density enhancements in the
channel corners) acts to only slightly reduce the diffusivity
relative to what would be predicted by the average density $\rho$.

\begin{figure}
\scalebox{.9}{!}{\includegraphics{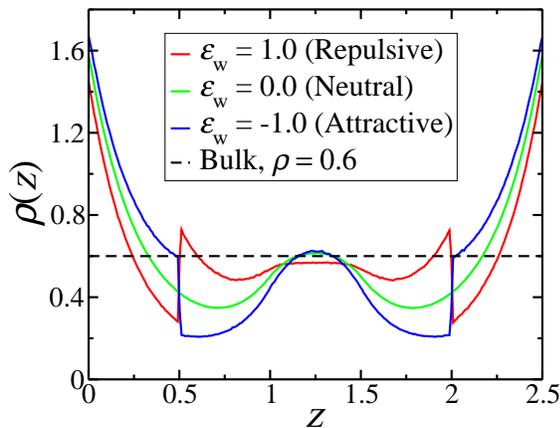}}
\caption{\label{profile}Density profiles $\rho(z)$ for HS fluids
confined to a 2$d$ channel of width 
$h_z=2.5$.  Although each system has different particle-wall 
interaction strengths
$\epsilon_{\mathrm w}$, they share a common average density $\rho$ (dashed) and
self-diffusivity $D$.}
\end{figure}

Given the approximate collapse of the data in
Fig.~\ref{D_density}, it is 
natural to wonder whether it is simply the particle structuring 
that is determining the HS dynamics.  
To test this idea further, we examine in Fig.~\ref{profile}
the local density profiles $\rho(z)$ for a HS fluid confined to 2$d$ channels
of width $h_z=2.5$ but with three different particle-wall
interactions: $\epsilon_{\mathrm w}=1$ (repulsive walls), 
$\epsilon_{\mathrm w}=0$ (neutral walls), and 
$\epsilon_{\mathrm w}=-1$ (attractive walls).  All three systems
exhibit the same average density $\rho$, and thus according to 
Fig.~\ref{D_density},  display approximately 
the same self-diffusivity $D$ as the bulk fluid.  Clearly there are
real and pronounced differences in the local structuring of the three confined 
fluids, especially when compared to the uniform bulk material.  These
types of structural differences are usually the main focus of
studies of inhomogenenous fluids by classical density functional
theories.  Interestingly, these pronounced structural differences
only slightly alter both $s^{\text{ex}}$ and $D$ when considering 
fluids with the same average $\rho$.

\begin{figure}[t]
\scalebox{.9}{!}{\includegraphics{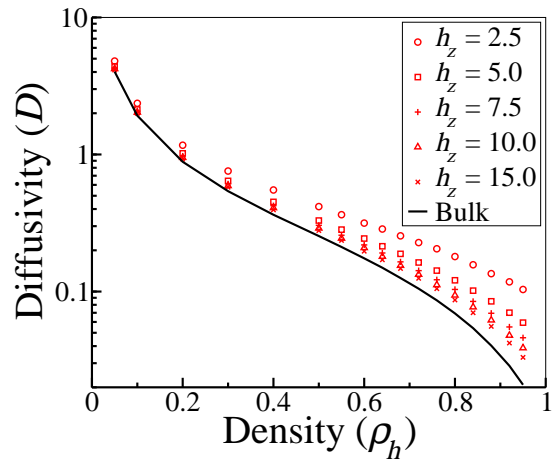}}
\caption{\label{D_rhoh} 
Self-diffusivity $D$ versus average density 
$\rho_h= N/V$ based on the the total particle-center-accessible
volume $V$. 
Systems shown include the bulk 
HS fluid (solid curve) and the HS fluid confined to 
the 2$d$ channels between hard walls.} 
\end{figure}

Finally, we examine what conclusions regarding density 
follow if one instead chooses to plot $D$ versus the alternative 
definition for 
average density, $\rho_h=N/V$, based on the total
particle-center-accessible volume $V$.  In particular,
Fig.~\ref{D_rhoh} compares the self-diffusivity for the HS 
fluid confined to 5 different 2$d$ channels with hard walls 
($\epsilon_{\mathrm w}=0$) as a function of $\rho_h$.  Unlike when
plotting versus $\rho$, there is no data collapse of $D$ when plotting
versus $\rho_h$.  Thus, one might consider $\rho$ a more natural
independent variable than $\rho_h$, not only for
thermodynamics of inhomogeneous fluids~\cite{lowen2dp,rice2dp}, 
but also for dynamics.

To conclude, we have probed the structure, entropy, and diffusivity of
the HS fluid confined to 2$d$ and 1$d$ channels with a wide
range of dimensions and particle-boundary interactions.  
Our main finding is that the
relationships between diffusivity, excess entropy, and 
average density
for the bulk HS fluid also remain valid, 
to within an excellent approximation,
for the HS fluid confined to particle-scale geometries.
Since statistical mechanical theories can provide accurate estimates for how
confinement modifies the excess entropy and density, the robust
connection between thermodynamics and dynamics reported here
should have far-reaching implications for the prediction of dynamics 
in confined systems.  We are currently testing
whether similar connections hold (i) for the HS fluid in more general random
environments (e.g., in quenched media, etc.), and (ii) for fluids
with strong interparticle attractions that significantly affect
local structuring.

TMT and JRE acknowledge the support of the National
Science Foundation Grants No. CTS-0448721 and CTS-028772,
respectively, and the Donors of the American Chemical Society
Petroleum Research Fund Grants No. 41432-G5 and 43452-AC5, respectively.
TMT acknowledges the support of a Packard Fellowship.

\bibliography{jjt}

\begin{thebibliography}{21}
\expandafter\ifx\csname natexlab\endcsname\relax\def\natexlab#1{#1}\fi
\expandafter\ifx\csname bibnamefont\endcsname\relax
  \def\bibnamefont#1{#1}\fi
\expandafter\ifx\csname bibfnamefont\endcsname\relax
  \def\bibfnamefont#1{#1}\fi
\expandafter\ifx\csname citenamefont\endcsname\relax
  \def\citenamefont#1{#1}\fi
\expandafter\ifx\csname url\endcsname\relax
  \def\url#1{\texttt{#1}}\fi
\expandafter\ifx\csname urlprefix\endcsname\relax\def\urlprefix{URL }\fi
\providecommand{\bibinfo}[2]{#2}
\providecommand{\eprint}[2][]{\url{#2}}

\bibitem[{\citenamefont{Reiter}(1994)}]{reiter}
\bibinfo{author}{\bibfnamefont{G.}~\bibnamefont{Reiter}},
  \bibinfo{journal}{Macromolecules} \textbf{\bibinfo{volume}{27}},
  \bibinfo{pages}{3046} (\bibinfo{year}{1994}).

\bibitem[{\citenamefont{Morineau et~al.}(2002)\citenamefont{Morineau, Xia, and
  Alba-Simionesco}}]{alba1}
\bibinfo{author}{\bibfnamefont{D.}~\bibnamefont{Morineau}},
  \bibinfo{author}{\bibfnamefont{Y.}~\bibnamefont{Xia}}, \bibnamefont{and}
  \bibinfo{author}{\bibfnamefont{C.}~\bibnamefont{Alba-Simionesco}},
  \bibinfo{journal}{J. \ Chem. \ Phys.} \textbf{\bibinfo{volume}{117}},
  \bibinfo{pages}{8966} (\bibinfo{year}{2002}).

\bibitem[{\citenamefont{McCoy and Curro}(2002)}]{curro}
\bibinfo{author}{\bibfnamefont{J.~D.} \bibnamefont{McCoy}} \bibnamefont{and}
  \bibinfo{author}{\bibfnamefont{J.~G.} \bibnamefont{Curro}},
  \bibinfo{journal}{J. \ Chem. \ Phys.} \textbf{\bibinfo{volume}{116}},
  \bibinfo{pages}{9154} (\bibinfo{year}{2002}).

\bibitem[{\citenamefont{Jones}(1999)}]{jones}
\bibinfo{author}{\bibfnamefont{R.~A.~L.} \bibnamefont{Jones}},
  \bibinfo{journal}{Curr. \ Opin. \ Colloid \ Interface \ Sci.}
  \textbf{\bibinfo{volume}{94}}, \bibinfo{pages}{167} (\bibinfo{year}{1999}).

\bibitem[{\citenamefont{Scheidler et~al.}(2004)\citenamefont{Scheidler, Kob,
  and Binder}}]{binder2}
\bibinfo{author}{\bibfnamefont{P.}~\bibnamefont{Scheidler}},
  \bibinfo{author}{\bibfnamefont{W.}~\bibnamefont{Kob}}, \bibnamefont{and}
  \bibinfo{author}{\bibfnamefont{K.}~\bibnamefont{Binder}},
  \bibinfo{journal}{J. \ Phys. \ Chem. \ B} \textbf{\bibinfo{volume}{108}},
  \bibinfo{pages}{6673} (\bibinfo{year}{2004}).

\bibitem[{\citenamefont{Fehr and Lowen}(1995)}]{lowen_ss}
\bibinfo{author}{\bibfnamefont{T.}~\bibnamefont{Fehr}} \bibnamefont{and}
  \bibinfo{author}{\bibfnamefont{H.}~\bibnamefont{Lowen}},
  \bibinfo{journal}{Phys. \ Rev. \ E} \textbf{\bibinfo{volume}{52}},
  \bibinfo{pages}{4016} (\bibinfo{year}{1995}).

\bibitem[{\citenamefont{Rosenfeld}(1977)}]{rosen1}
\bibinfo{author}{\bibfnamefont{Y.}~\bibnamefont{Rosenfeld}},
  \bibinfo{journal}{Phys. \ Rev. \ A} \textbf{\bibinfo{volume}{15}},
  \bibinfo{pages}{2545} (\bibinfo{year}{1977}).

\bibitem[{\citenamefont{Dzugutov}(1996)}]{dzu1}
\bibinfo{author}{\bibfnamefont{M.}~\bibnamefont{Dzugutov}},
  \bibinfo{journal}{Nature} \textbf{\bibinfo{volume}{381}},
  \bibinfo{pages}{137} (\bibinfo{year}{1996}).

\bibitem[{\citenamefont{Rosenfeld}(1999)}]{rosen2}
\bibinfo{author}{\bibfnamefont{Y.}~\bibnamefont{Rosenfeld}},
  \bibinfo{journal}{J. \ Phys.: \ Condens. \ Matter}
  \textbf{\bibinfo{volume}{11}}, \bibinfo{pages}{5415} (\bibinfo{year}{1999}).

\bibitem[{\citenamefont{Bretonnet}(2002)}]{bret}
\bibinfo{author}{\bibfnamefont{J.~L.} \bibnamefont{Bretonnet}},
  \bibinfo{journal}{J. \ Chem. \ Phys.} \textbf{\bibinfo{volume}{117}},
  \bibinfo{pages}{9370} (\bibinfo{year}{2002}).

\bibitem[{\citenamefont{Bastea}(2003)}]{bast}
\bibinfo{author}{\bibfnamefont{S.}~\bibnamefont{Bastea}},
  \bibinfo{journal}{Phys. \ Rev. \ E} \textbf{\bibinfo{volume}{68}},
  \bibinfo{pages}{031204} (\bibinfo{year}{2003}).

\bibitem[{\citenamefont{Samanta et~al.}(2001)\citenamefont{Samanta, Ali, and
  Ghosh}}]{sam1}
\bibinfo{author}{\bibfnamefont{A.}~\bibnamefont{Samanta}},
  \bibinfo{author}{\bibfnamefont{S.~M.} \bibnamefont{Ali}}, \bibnamefont{and}
  \bibinfo{author}{\bibfnamefont{S.~K.} \bibnamefont{Ghosh}},
  \bibinfo{journal}{Phys. \ Rev. \ Lett.} \textbf{\bibinfo{volume}{87}},
  \bibinfo{pages}{245901} (\bibinfo{year}{2001}).

\bibitem[{\citenamefont{Samanta et~al.}(2004)\citenamefont{Samanta, Ali, and
  Ghosh}}]{sam2}
\bibinfo{author}{\bibfnamefont{A.}~\bibnamefont{Samanta}},
  \bibinfo{author}{\bibfnamefont{S.~M.} \bibnamefont{Ali}}, \bibnamefont{and}
  \bibinfo{author}{\bibfnamefont{S.~K.} \bibnamefont{Ghosh}},
  \bibinfo{journal}{Phys. \ Rev. \ Lett.} \textbf{\bibinfo{volume}{92}},
  \bibinfo{pages}{145901} (\bibinfo{year}{2004}).

\bibitem[{\citenamefont{Hoyt et~al.}(2000)\citenamefont{Hoyt, Asta, and
  Sadigh}}]{hoyt}
\bibinfo{author}{\bibfnamefont{J.~J.} \bibnamefont{Hoyt}},
  \bibinfo{author}{\bibfnamefont{M.}~\bibnamefont{Asta}}, \bibnamefont{and}
  \bibinfo{author}{\bibfnamefont{B.}~\bibnamefont{Sadigh}},
  \bibinfo{journal}{Phys. \ Rev. \ Lett.} \textbf{\bibinfo{volume}{85}},
  \bibinfo{pages}{594} (\bibinfo{year}{2000}).

\bibitem[{\citenamefont{Rapaport}(2004)}]{rap}
\bibinfo{author}{\bibfnamefont{D.~C.} \bibnamefont{Rapaport}},
  \emph{\bibinfo{title}{The Art of Molecular Dynamics Simulation}}
  (\bibinfo{publisher}{Cambridge University Press}, \bibinfo{year}{2004}),
  \bibinfo{edition}{2nd} ed.

\bibitem[{\citenamefont{Errington}(2003)}]{jeff1}
\bibinfo{author}{\bibfnamefont{J.~R.} \bibnamefont{Errington}},
  \bibinfo{journal}{J. \ Chem. \ Phys.} \textbf{\bibinfo{volume}{118}},
  \bibinfo{pages}{9915} (\bibinfo{year}{2003}).

\bibitem[{not()}]{note1}
\bibinfo{note}{The activity is defined as $\xi=\exp(\beta \mu)/\Lambda^3$,
  where $\mu$ is the chemical potential and $\Lambda$ is the de Broglie
  wavelength.}

\bibitem[{\citenamefont{Panagiotopoulos}(2000)}]{azp00}
\bibinfo{author}{\bibfnamefont{A.~Z.} \bibnamefont{Panagiotopoulos}},
  \bibinfo{journal}{J. Phys.: Condens. Matter} \textbf{\bibinfo{volume}{12}},
  \bibinfo{pages}{R25} (\bibinfo{year}{2000}).

\bibitem[{\citenamefont{Davis}(1996)}]{htdavis}
\bibinfo{author}{\bibfnamefont{H.~T.} \bibnamefont{Davis}},
  \emph{\bibinfo{title}{Statistical Mechanics of Phases, Interfaces, and Thin
  Films}} (\bibinfo{publisher}{VCH}, \bibinfo{year}{1996}).

\bibitem[{\citenamefont{Schmidt and L{\"{o}}wen}(1997)}]{lowen2dp}
\bibinfo{author}{\bibfnamefont{M.}~\bibnamefont{Schmidt}} \bibnamefont{and}
  \bibinfo{author}{\bibfnamefont{H.}~\bibnamefont{L{\"{o}}wen}},
  \bibinfo{journal}{Phys. \ Rev. \ E} \textbf{\bibinfo{volume}{55}},
  \bibinfo{pages}{7228} (\bibinfo{year}{1997}).

\bibitem[{\citenamefont{Zangi and Rice}(1998)}]{rice2dp}
\bibinfo{author}{\bibfnamefont{R.}~\bibnamefont{Zangi}} \bibnamefont{and}
  \bibinfo{author}{\bibfnamefont{S.~A.} \bibnamefont{Rice}},
  \bibinfo{journal}{Phys. \ Rev. \ E} \textbf{\bibinfo{volume}{58}},
  \bibinfo{pages}{7529} (\bibinfo{year}{1998}).

\end{thebibliography}

\end{document}